\documentclass[structabstract]{aa}
\usepackage{natbib}
\bibpunct{(}{)}{;}{a}{}{,}

\usepackage{graphicx}
\usepackage{txfonts}
\usepackage{color}

\def\beq{\begin{equation}}
\def\eeq{\end{equation}}
\def\bea{\begin{eqnarray}}
\def\eea{\end{eqnarray}}
\def\eqref#1{Eq.~(\ref{eq:#1})}
\def\eqlab#1{\label{eq:#1}}

\newcommand*{\figref}[1]{Fig.~\ref{fig:#1}}
\newcommand*{\figlab}[1]{\label{fig:#1}}
\newcommand*{\appref}[1]{Appendix~\ref{sec:#1}}
\newcommand*{\secref}[1]{Section~\ref{sec:#1}}
\newcommand*{\seclab}[1]{\label{sec:#1}}

\newcommand{\Omit}[1]{}

\begin{document}
   \title{Constraints on the flux of Ultra-High Energy neutrinos from WSRT observations}
   \titlerunning{UHE neutrino limits}

\author{S. Buitink \inst{1,2}
 O. Scholten  \inst{3}
 J. Bacelar \inst{4}
 R. Braun \inst{5}
 A.G. de Bruyn \inst{6,7}
 H. Falcke \inst{2,7}
 K. Singh \inst{3}
 B. Stappers \inst{8}
 R.G. Strom \inst{7,9}
 R. al Yahyaoui \inst{3}}
\authorrunning{Buitink et al.}

\institute{
$^1$Lawrence Berkeley National Laboratory, Berkeley, CA 94720, USA\\
$^2$Department of Astrophysics, IMAPP, Radboud University, 6500 GL Nijmegen, The
Netherlands\\
$^3$Kernfysisch Versneller Instituut, University of Groningen, 9747
AA, Groningen, The Netherlands\\
$^4$ASML Netherlands BV, P.O.Box 324, 5500 AH Veldhoven, The Netherlands\\
$^5$CSIRO - Astronomy and Space Science, P.O.Box 76, Epping NSW 1710, Australia\\
$^6$Kapteyn Institute, University of Groningen, 9747 AA, Groningen, The
Netherlands\\
$^7$ASTRON, 7990 AA Dwingeloo, The Netherlands\\
$^8$School of Physics \&
Astronomy, Alan Turing Building, Univ. of Manchester,
Manchester, M13 9PL\\
$^9$Astronomical Institute `A. Pannekoek', University of Amsterdam, 1098 SJ, The
Netherlands }

   \date{Received -; accepted -}

 \abstract{Ultra-high energy (UHE) neutrinos and cosmic rays initiate particle cascades underneath the Moon's surface. These cascades have a
 negative charge excess and radiate Cherenkov radio emission in a process known as the Askaryan effect. The optimal frequency window for
 observation of these pulses with radio telescopes on the Earth is around 150 MHz.
 }{By observing the Moon with the Westerbork Synthesis Radio Telescope array we are able to set a new limit on the UHE neutrino
 flux.
 }{The PuMa II backend is used to monitor the Moon in 4 frequency bands between 113 and 175 MHz with a sampling frequency of 40 MHz. The
 narrowband radio interference is digitally filtered out and the dispersive effect of the Earth's ionosphere is compensated for. A
 trigger system is implemented to search for short pulses. By inserting simulated pulses in the raw data, the detection efficiency for pulses
 of various strength is calculated.
 }{With 47.6 hours of observation time, we are able to set a limit on the UHE neutrino flux. This new limit is an order of magnitude
 lower than existing limits. In the near future, the digital radio array LOFAR will be used to achieve an even lower limit.
 }
 {}


   \maketitle
%

\section{introduction}

The cosmic ray energy spectrum follows a power law distribution extending up to extremely large energies. At the Pierre Auger Observatory (PAO) cosmic rays (CRs) are observed up
to energies around $\sim 10^{20}$~eV.
Above the Greisen-Zatsepin-Kuzmin (GZK) energy of $6\cdot 10^{19}$~eV, CRs can interact with the cosmic
microwave background photons.
In the most efficient interaction, a $\Delta$-resonance is produced which decays into either a proton and a
neutral pion or a neutron and a positively charged pion. Charged pions decay and produce neutrinos. The energy loss length for
$\Delta$-resonance production is $\sim 50$\, Mpc \citep{G66,ZK66}.

Recent results of the PAO have confirmed a steepening in the cosmic ray spectrum
at the GZK energy \citep{A08}. This steepening is not necessarily a clear
cut-off, as CRs from local sources may arrive at Earth with super-GZK
energies. Because of their large energies, these particles will only deflect
slightly in the (extra-) Galactic field during their propagation, and their
arrival directions correlate with their sources \citep{A07}.

Sources at distances larger than 50 Mpc can be found by observing neutrinos that are produced in GZK interactions.
Since neutrinos are chargeless they will propagate in a straight line from the location where the GZK interaction took place to the observer, thus conserving the directional
information.
In addition, while CRs from distant sources pile up at the GZK energy, information
about the CR spectrum at the source is conserved in the GZK neutrino flux.
Other possible sources of UHE neutrinos are decaying supermassive dark matter particles or
topological defects. This class of models is refered to as top-down models (see for example \citet{s04} for
a review).

Because of their small interaction cross section and low flux, the detection of
cosmic neutrinos calls for extremely large detectors. Assuming the Waxman-Bahcall
flux \citep{wb98,wb01}, even at low energies in the GeV range, the flux is not
higher than a few tens of neutrinos per km$^2$ per year. Kilometer-scale
detectors are not easily built but can be found in nature. For example,
interaction of neutrinos in ice or water can be detected by the Cherenkov light
produced by the lepton track or cascade. The nearly completed IceCube
detector \citep{icecube} will cover a km$^3$ volume of South Pole ice with
optical modules, while Antares \citep{ antares} and its successor KM3NET
\citep{KM3NET} exploit the same technique in the Mediterranean sea. Even larger
volumes can be covered by observing large detector masses from a distance. The
ANITA balloon mission \citep{anita} monitors an area of a million km$^2$ of South
Pole ice from an altitude of $\sim 37$~km and the FORTE satellite \citep{forte}
can pick up radio signals coming from the Greenland ice mass. Alternatively,
cosmic ray experiments like the
 Pierre Auger
Observatory can possibly distinguish cosmic ray induced air showers from neutrino induced cascades at very
high zenith angles where the atmosphere is thickest and only neutrinos can interact close to the detector.

The Moon offers an even larger natural detector volume. When CRs or neutrinos hit the Moon they will interact with the medium. CRs will start a particle cascade just below the
Lunar surface, while neutrinos will interact deeper inside the Moon, also creating a hadronic shower. The negative charge excess of a particle cascade inside
a dense medium will cause the emission of coherent Cherenkov radiation in a process known as the Askaryan
effect \citep{a62}. This emission mechanism has been experimentally verified at accelerators \citep{s01,g00} and extensive
calculations have been
performed to quantify the effect \citep{zhs92,az97}. The idea to observe this type of emission from the Moon
with radio telescopes was first proposed by \citet{dz89} and the first experimental endeavours in this
direction were carried out with the Parkes telescope \citep{parkes}, at Goldstone (GLUE) \citep{glue}, and with the Kalyazin Radio Telescope \citep{KALYAZIN}.
LUNASKA (Lunar UHE Neutrino Astrphysics with the Square Kilometer Array) is a
project that is currently performing lunar Cherenkov measurements with ATCA (the
Australia Telescope Compact Array) with a 600 MHz bandwidth at 1.2-1.8 GHz \citep{LUNASKA}.

\citet{fg03} suggested to use low-frequency telescopes (like LOFAR) for such an
experiment. It is shown by \citet{scholten} that observing at lower frequencies
has the distinct advantage that radio pulses have a much higher chance of
reaching the observer, as will be explained in the next section. In this work we
use data recorded with the Westerbork Synthesis Radio Telescope (WSRT) in the
frequency range of 113-168~MHz to set a new limit on the flux of UHE
neutrinos. A first reporting of this limit is made in \citep{NuMoon-PRL}.

\section{Detection principle}
UHE neutrinos or CRs interact below the lunar surface. In the case of a CR, all energy is converted into a hadronic shower. In a neutrino interaction, only about 20\% of
the energy is converted into a hadronic shower, while the other 80\% is carried off by a lepton (corresponding to the neutrino flavor), which will not produce
any observable radio emission. Muons will not produce
enough charge density, while electromagnetic showers become elongated at energies above $E_{LPM}=10^{18}$~eV due to the
Landau-Pomeranchuk-Migdal (LPM) effect \citep{LPM}. For these showers the angular spread of the radio emission around the Cherenkov
angle becomes very small, severely lowering the chance of detection.

For proton energies exceeding $10^{20}$~eV it is predicted that the LPM effect will start to play a role since many of the
leptons and photons which are created as secondary particles have energies in
excess of the $E_{LPM}$. This has the effect of
creating a lopsided hadronic shower with a rather long 'tail' \citep{LPM}. The
bulk of the charged particles in the shower is
still present over a length which
one would have obtained ignoring the LPM effect and our estimates should thus
apply also to the hadronic part of showers initiated by neutrinos of energies
ranging up to $10^{23}$~eV.

The lateral size of the cascade is of the order of 10~cm so the radio emission is coherent up to $\sim 3$ GHz. Former
experiments, like GLUE, have observed at high frequencies (2.2~GHz) where the emission is strongest. For lower frequencies, however, the
angular spread of the emission around the Cherenkov angle increases due to diffraction. For
emission at the Cherenkov angle, only those showers can be observed that hit the rim of the Moon, under such an angle that the emission will not be internally
reflected at the Lunar surface. With a larger angular spread in the emission a wider range of
geometries is allowed and a larger part of the lunar surface acts as a radiation source. When the wavelength is of the order of the shower
length, several meters, the emission becomes nearly isotropic and pulses can be expected to come from the whole Moon \citep{scholten}.
In our experiment we exploit this optimal frequency range around 150 MHz.

The intensity of the radio emission from a hadronic shower with energy $E_s$ in the lunar regolith can be parameterized
as \citep{zhs92,az97,scholten}
\begin{eqnarray}
F(\theta,\nu,E_s)=3.86\cdot 10^4 \,e^{-Z^2} \left(\frac{\sin\theta}{\sin\theta_c}\right)^2
\left(\frac{E_s}{10^{20}\mathrm{eV}}\right)^2
\left(\frac{d_{\mathrm{moon}}}{d}\right)^2 \nonumber \\
\left(\frac{\nu}{\nu_0 (1+(\nu/\nu_0)^{1.44})}\right)^2
\left(\frac{\Delta \nu}{\mathrm{100 MHz}}\right) \mathrm{Jy},
\end{eqnarray}
with
\begin{equation}
Z=(\cos\theta-1/n)\left(\frac{n}{\sqrt{n^2-1}}\right)\left(\frac{180}{\pi\Delta_c}\right),
\end{equation}
where $\Delta \nu$ is the bandwidth, $\nu$ the central frequency and $\nu_0=2.5$~GHz. The average Earth-Moon distance
$d_\mathrm{moon}=3.844\cdot 10^{8}$~m, $d$ is the distance to the observer. The Cherenkov angle is given by
$\cos\theta_c=1/n$, where $n$ is the index of refraction and $\theta$ is the angle under which radiation is emitted
relative to the direction of shower propagation. The spread of radiation around the Cherenkov angle is given by
\begin{equation}
\Delta_c=4.32^{\circ}\left(\frac{1}{\nu[\mathrm{GHz}]}\right)\left(\frac{L(10^{20}\mathrm{eV})}{L(E_s)}\right),
\end{equation}
where $L$ is the shower length depending on primary energy.

The regolith is the top layer of the Moon and consists of dust and small rocks.
The properties of this layer are known from samples brought from the Moon by the
Apollo missions \citep{os75}. The average index of refraction is $n=1.8$ and the
mean attenuation length is found to be $\lambda_{r}=(9/\nu[\mathrm{GHz}])$~m
for radio waves \citep{os75,hvf91}. There are sizable differences in,
especially, the reported values of the attenuation length. The effects of this
uncertainty on the extracted limits is discussed in \secref{results}. The
thickness of the regolith is known to vary over the lunar surface. At some depth
there is a (probably smooth) transition to solid rock, for which the density is
about twice that of the regolith. \citet{wz01} report that the density is almost
homogeneous down to a depth of 20~km. In \citet{scholten} the effects of pure
rock and regolith are simulated and found to give very similar detection limits
for low frequencies.

As the radiation leaves the Moon it refracts through the surface. In
\citet{glue} and \citet{james} the effects of this refraction for smooth and
irregular surfaces are described. It is shown that the angular spread $\Delta
\theta$ increases due to this refraction and that this effect is especially
strong when the angle at which the radiation approaches the lunar surface is
close to the angle of total internal reflection. The larger angular spread
increases the acceptance but also increases the energy threshold for detection
since the radiated power spreads out over a larger area. Small scale
irregularities of the lunar surface make this effect stronger because variations
in the surface tangent within the radiation beam will increase the angular spread
by refraction even more. At the frequencies at which we observe, these effects
are of less importance since $\Delta \theta$ is already large at the source due
to diffraction and thus the increase due to surface irregularities can safely
be ignored.

\section{Detection with WSRT}
\label{sec:detection}
The Westerbork Radio Synthesis Telecope (WSRT) is an array telescope consisting of 14 parabolic
telescopes of 25~m on a 2.7~km east-west line. The NuMoon experiment uses the Low Frequency Front Ends (LFFEs)
which cover the frequency range 115--180 MHz. Each LFFE records full polarization data.
For our observations we use the Pulsar Machine II (PuMa II)
backend \citep{kss08}, which can record a maximum bandwidth of 160 MHz, sampled as 8 subbands of 20 MHz each.

Only 11 of the 12 equally spaced WSRT dishes are used for this experiment
which means that when the telescopes are added in phase the resultant beam on the
sky is a fan beam \citep{J09}. The phases required to add the dishes coherently
are determined by observations of a known calibrator source, which at these
frequencies is Cassiopeia A. Adjusting the phase relations between the 8 subbands
they can be pointed to any location within the primary beam of the 25 m dish.

We use
two beams of 4 bands each, centered around 123, 137, 151, and 165 MHz. The two beams are aimed at different
sides of the Moon, each covering about one third of the lunar surface, in order to
enlarge the effective aperture and create the possibility of an anti-coincidence trigger. A lunar Cherenkov pulse should only be visible in one of the two beams. Because of overlap in the subbands the
total bandwidth per beam is 65 MHz. The system has a real time automatic gain control (AGC) system, that stabilizes the average gain of the output signal.

For each subband, the time series data is recorded at several storage nodes with a sampling frequency
of 40 MHz.

The data is processed in blocks of 0.1 s, each block being divided in 200 traces of
20\ 000 time samples. The signal of individual WSRT dishes is 2 bit, limiting the
dynamic range of an 11-dish observation to 34. We will discuss the implications
of this limited dynamic range in Sec.\ \ref{sec:simulations}. There is data for
two beams, each containing 4 frequency bands and 2 polarization directions.

The data analysis is performed in the following steps:
\begin{itemize}
\item{\bf RFI background reduction} Radio Frequency Interference (RFI) is
    narrow band anthropogenic emission, which can be responsible for a large
    part of the received power and has to be filtered out of the data. For
    all time traces an FFT is produced and for each data block the 200
    frequency spectra are added to obtain an integrated frequency spectrum.
    The baseline of this spectrum is fitted with a $9^{th}$ order
    polynomial function and bins containing a value exceeding the fit by 50\%
    are marked as RFI lines. In each individual frequency spectrum all bins
    that are marked as RFI lines are set to zero. This procedure is carried
    out separately for each of the 4 frequency bands and the 2
    polarizations. The number of RFI lines per spectrum varies with time and
    is different for all frequency bands and polarizations, but does seldom
    exceed 200. The corresponding loss in bandwidth is $\sim$2\% at maximum.
    Figure \ref{RFI} shows frequency spectra of 10 seconds of data before RFI
    removal. In the highest frequency band the upper end of the spectrum is
    suppressed by a band pass filter, lowering the effective bandwidth. In
    other frequency bands a similar suppression can be seen, but this is
    compensated by the overlap between the different bands. Our effective
    bandwidth is 55 MHz, ranging from 113 to 168 MHz. An example of a ten
    second frequency spectrum after RFI removal is shown in
    Fig.~\ref{RFIremoved}.

\begin{figure}
\centering
\includegraphics[width=9cm]{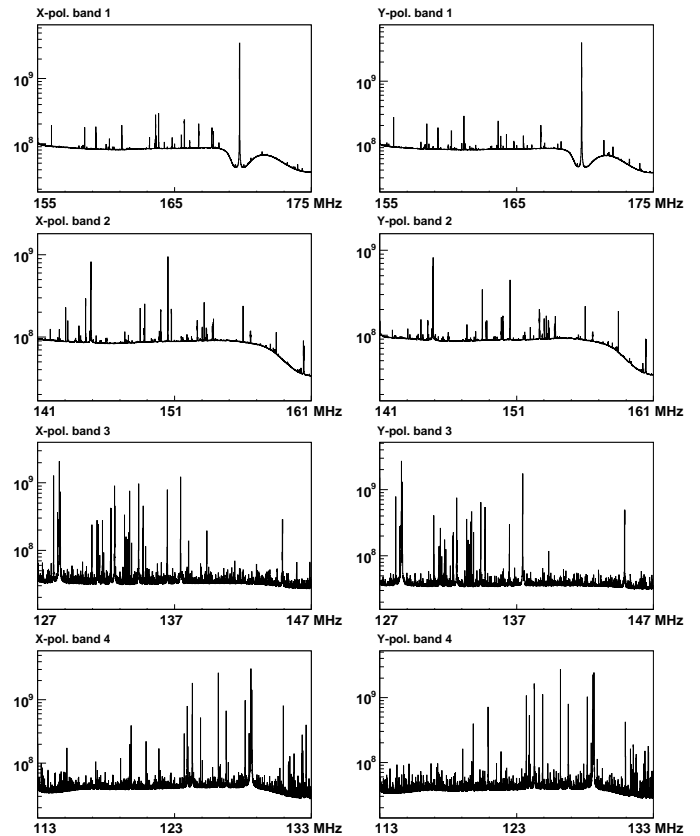}
\caption{Frequency spectra of 10 seconds of data for all bands and polarizations. The narrow RFI lines that exceed a fit to the
curve by 50\% are put to zero.}
\label{RFI}
\end{figure}

\begin{figure}
\centering
\includegraphics[width=7cm]{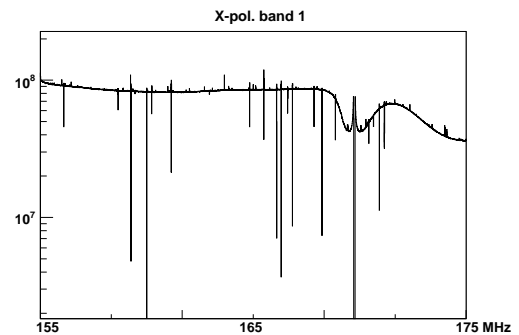}
\caption{Frequency spectra of 10 seconds of data for band 1 and $x$ polarization after RFI removal. The spectrum is composed of 100 spectra that have been subjected to RFI
removal separately. In each spectrum a frequency sample is set to zero when the amplitude exceeds the fitted background curve by more than 50\%. When a certain sample contains
an RFI line in all 100 spectra, it has value zero in this integrated spectra. Samples that have a non-zero value below the background curve correspond to RFI lines that are only
present in part of the 100 spectra.}
\label{RFIremoved}
\end{figure}

\item{\bf Ionospheric de-dispersion} After the RFI removal the data is still
    in the frequency domain. The de-dispersion is performed by applying a
    frequency dependent phase shift to the data before transforming back to
    the time domain. The Vertical TEC values, that are needed for the
    de-dispersion, are provided by the DLR Institut f\"ur Kommunikation und
    Navigation\footnote{http://www.dlr.de/kn}. These were used to
    calculate the STEC by compensating for the Moon elevation. Because of
    variations in the thickness of the ionosphere on short timescales and
    over distance, we assume the presence of an error in the de-dispersion,
    resulting in an increased time width of the pulses and an offset between
    the arrival times of pulses in different frequency bands. The
    implications hereof for our analysis are further discussed in Sec.\
    \ref{sec:simulations}.

\item{\bf Evaluation of $P_5$} After de-dispersion, an inverse FFT is performed to
transform the data back into the time domain. The cutting of RFI lines
increases the noise level in the time samples close to the edges of the time trace. The
de-dispersion can move this increased noise further backward in time. To avoid
triggering on this noise the first and last 250 time samples are excluded from
analysis, corresponding to 0.25\% of the observation time. Next, we
calculate $P_5$, the power integrated over 5 consecutive samples normalized over
one trace
\begin{equation}
P_5=\frac{\displaystyle\sum_{\mathrm{5\ samples}} P_x}{\bigg<\displaystyle\sum_{\mathrm{5\ samples}} P_x\bigg>}+
\frac{\displaystyle\sum_{\mathrm{5\ samples}} P_y}{\bigg<\displaystyle\sum_{\mathrm{5\ samples}} P_y\bigg>},
\label{P5}
\end{equation}
where the averaging is done over one time trace (20\,000 time samples), and $x$ and $y$ denote the two polarizations.
The integration has been chosen to be over 5 samples, because this is the typical
number of samples over which the power is spread for a bandwidth limited and Nyquist sampled
pulse with a small dispersion (see Appendix \ref{app:pulses}).

\item{\bf Pulse search} The data is scanned for values of $P_5$ exceeding 5. The meaning of this threshold can be understood from Eqn.\ \ref{P5}. If we define $\sigma^2$ as
the mean power in one time sample (assuming for simplicity that it is equal for the $x$- and $y$-polarizations), then the trigger condition can be written as
\begin{equation}
\displaystyle\sum_{\mathrm{5\ samples}} P_x + \displaystyle\sum_{\mathrm{5\ samples}} P_y > 25 \sigma^2,
\end{equation}
meaning that the total power in ten bins (five in the $x$ polarization and five in the $y$ polarization) must add up to a value larger than 25$\sigma^2$, where the average is 10$\sigma^2$.
The band with the highest frequency is first scanned for $P_5$ values
exceeding 5. When such a value is found the $P_5$ values of the other 3
frequency bands are evaluated near this position. A time
offset between pulses in the different bands of
\begin{equation}
\Delta t = 1.34 \cdot 10^{9} \cdot 0.30 \cdot \mathrm{STEC} \left( \frac{1}{\nu_{1}^2}-\frac{1}{\nu_{2}^2}\right)
\end{equation}
is allowed based on an error of 30\% on the STEC value. When a $P_5$ value exceeding 5 is found in all bands the time
trace is permanently stored, together with information on the RFI lines and the
data of the corresponding time trace in the other beam. No search is done for a
second pulse in the same trace. An estimation of the resulting loss in effective
observation time is given in Sec.\ \ref{sec:background}. For each trigger the
location, maximum value, width and offsets between locations in the different
bands are stored. The width is defined as the number of consecutive $P_5$
values that exceed 5. The value $S$ is defined as the sum over the maximum $P_5$ values in the 4 frequency bands
\begin{equation}
S=\sum_{\mathrm{4\ bands}} P_5 \;. \eqlab{S}
\end{equation}
\end{itemize}

\begin{figure}
\centering
\includegraphics[width=8cm]{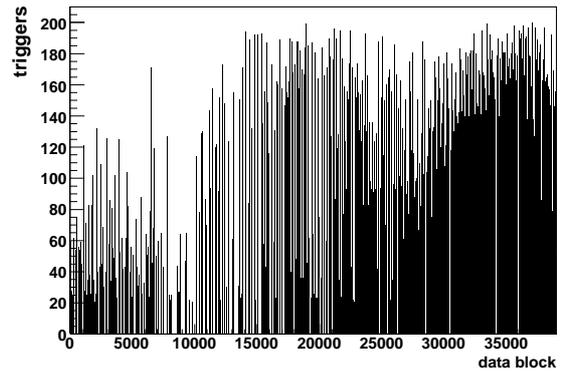}
\includegraphics[width=8cm]{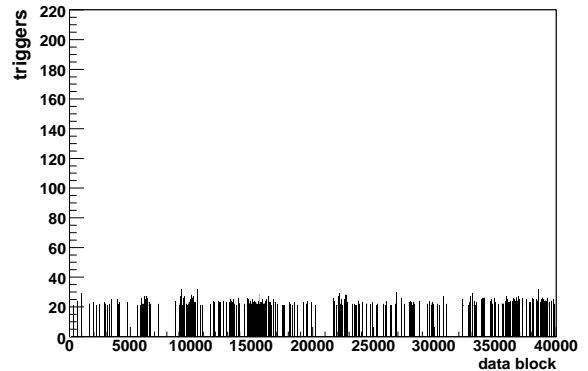}
\caption{Number of triggers per block is plotted against the number of the data block for one hour of data of the June 7, 2008
observation (top) and for one hour of the second observations of
August 29, 2008 (bottom). Only values exceeding 20 are plotted. The maximum value of 200 indicates completely saturated data.}
\label{june}
\end{figure}

\subsection*{Observations}
Table \ref{tab:obs} contains a list of the 22 observation runs performed between June 9, 2007 and November 11, 2008.
The number of raw triggers per hour is much higher than usual for the runs of June 9, 2007 and June 7/8, 2008.
The top panel in Figure \ref{june} shows the number of triggers per data block for one hour of one of
these runs. The number of triggers is only plotted when it exceeds 20. The maximum number of triggers per block is equal to the number of
traces: 200. This maximum is often reached, which means the run is not reliable.
The bottom panel shows the number of triggers per block for one hour of data of a regular observation
period. The observation runs for which the number of triggers per data block is exceptionally high
for a long period of time have been excluded from the analysis.

\begin{table*}
\caption[]{Observation runs}
\label{tab:obs}

\begin{tabular}{lrrrrr}
\hline
Date & Hours & STEC (lo/hi) & No.\ raw triggers & No.\ triggers after cuts ($S>$23)& No.\ triggers Gaussian noise ($S>$23) \\
\hline
2007 Jun 9$^{\mathrm{a}}$ & 4.7$^{\mathrm{b}}$  & 11.8/16.6 	& 200\ 427	& 49\ 679 	& 8\ 943 \\
2007 Sep 21 	 & 2 				& 15.6/19.5	& 668\ 917	& 26\ 812 	& 13\ 128 \\
2008 Jan 13 	 & 1.3$^{\mathrm{b}}$ 		& 18.0/24.3	& 119\ 032	& 6\ 951 	& 6\ 001 \\
2008 Jun 7$^{\mathrm{a}}$ & 4.25 		& 11.7/17.5	& 1\ 961\ 907	& 170\ 672	& 21\ 752 \\
2008 Jun 8$^{\mathrm{a}}$ & 5 			& 9.8/11.3	& 1\ 313\ 378	& 80\ 140	& 12\ 815 \\
2008 Aug 24 	 & 5 				& 3.5/7.5	& 792\ 979	& 36\ 314 	& 4\ 029 \\
2008 Aug 29 	 & 3 				& 6.5/7.0	& 563\ 692	& 45\ 214	& 3\ 331 \\
2008 Aug 29	 & 2 				& 8.0/8.3	& 602\ 049	& 29\ 554	& 4\ 317 \\
2008 Aug 29$^{\mathrm{a}}$  	 & 4.8 		& 6.5/9.7	& 1\ 719\ 443	& 96\ 998	& 7\ 425 \\
2008 Sep 2 	 & 5 				& 12.8/15.3	& 880\ 508	& 51\ 329	& 18\ 937 \\
2008 Sep 16	 & 3.75 			& 5.5/11.0	& 233\ 192	& 23\ 733	& 6\ 616 \\
2008 Sep 16	 & 5 				& 5.9/7.2	& 163\ 819	& 20\ 138	& 4\ 841 \\
2008 Sep 21 	 & 4.5 				& 3.3/4.8	& 244\ 276	& 27\ 388	& 3\ 451 \\
2008 Sep 21$^{\mathrm{a}}$	 & 5 		& 4.0/12.8	& 1\ 282\ 457	& 76\ 573	& 9\ 728 \\
2008 Sep 28	 & 3.7 				& 10.5/12	& 65\ 725	& 67\ 910	& 10\ 345 \\
2008 Sep 28	 & 3.8 				& 11.6/13.7	& 622\ 598	& 47\ 580	& 11\ 958 \\
2008 Oct 14	 & 4.5				& 5.9/9.9	& 566\ 611	& 58\ 127	& 6\ 531 \\
2008 Oct 14	 & 4.5				& 5.9/7.4	& 217\ 113	& 30\ 346	& 4\ 165 \\
2008 Nov 11	 & 3.7				& 3.5/7.4	& 941\ 369	& 27\ 160	& 2\ 624 \\

\hline
Total$^{\mathrm{c}}$ & 51.1 & 3.3/24.3 & 6\ 681\ 880& 430\ 646 & 100\ 274 \\ 
\hline
\end{tabular}
\begin{list}{}{}
\item[$^{\mathrm{a}}$] Excluded from analysis due to exceptional amount of raw triggers.
\item[$^{\mathrm{b}}$] Only single beam data available.
\item[$^{\mathrm{c}}$] Not counting excluded runs.
\end{list}
\end{table*}

\begin{figure}
\centering
\includegraphics[width=9cm]{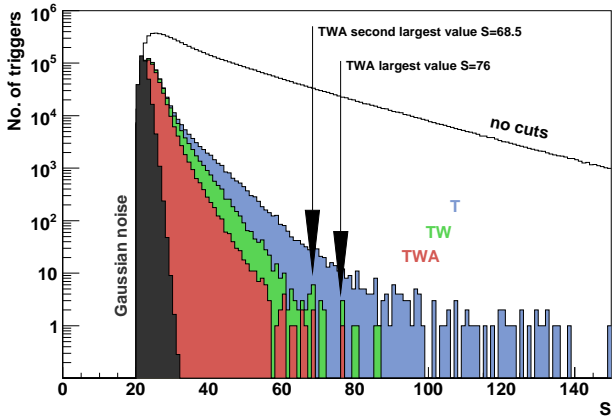}
\caption{Distribution of $S$ for several cuts. The top curve represents the distribution of raw
triggers. The other distribution represent, in order of decreasing number of contained events, the T, TW, and TWA cut. The black area corresponds to the distribution of triggers that are expected for a background of pure Gaussian noise.}
\label{distribution}
\end{figure}

Figure \ref{distribution} shows the distribution of $S$ for the triggered events.
The top curve corresponds to the raw triggers. Several additional cuts are applied:
\begin{itemize}
\item{\bf Timer signal (T)}
The data contains short strong pulses that repeat at a regular interval. They
can be visualized by plotting a distribution of pulse times folded by an appropriate time interval. Figure \ref{timer} shows the number of triggers
in 10 seconds of data against the number of the
time sample folded by 390,625. This corresponds to a frequency of 102.4 s$^{-1}$.

The specific time interval of these pulses suggest a technical origin. Cutting out the time intervals in which these pulses
occur corresponds to a loss of $\sim$10\% of observation time.

\begin{figure}
\centering
\includegraphics[width=9cm]{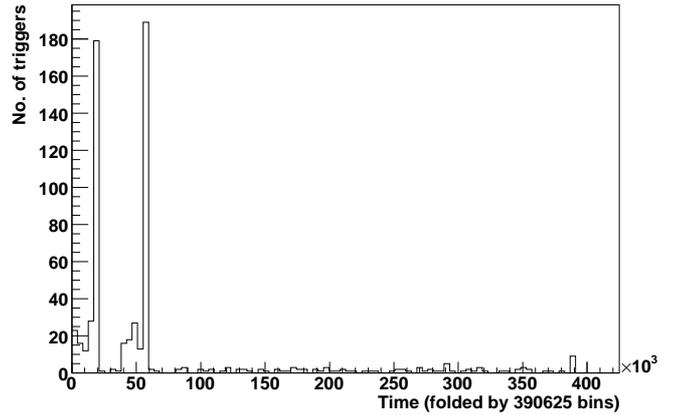}
\caption{Number of triggers vs.\ time sample folded by 390,625 samples. With this folding many triggers occur at the same time, probably having
a local technical origin. Triggers that occur at the positions of the peaks are excluded in the timer cut.}
\label{timer}
\end{figure}

\begin{figure}
\centering
\includegraphics[width=9cm]{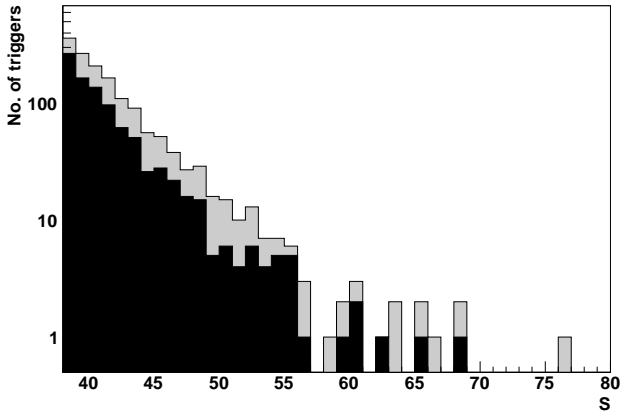}
\caption{Distribution of highest values for $S$ corresponding to a cut $W<10$ (dark) and $W<12$ (light). Cut on timer and anti-coincidence is applied in both cases.}
\figlab{twowidths}
\end{figure}

\item{\bf $P_5$ width (W)} We define the width of a pulse $W$ as the
    number of consecutive $P_5$ values exceeding the threshold.  For a real
    lunar pulse, $W$ should be limited. However, for increasingly tighter
    cuts on $W$, the probability of excluding a proper pulse grows. The
    value for the cut on $W$ was determined by examining the efficiency for
    recovering pulses in a simulation, as is explained in
    \secref{simulations}, showing a recovery rate of over 80\% by choosing
    $W<12$ for all four frequency bands. Since this value is not obtained by
    optimizing the distribution of $S$ we have avoided to introduce a bias by
    following this procedure.

\item{\bf Anti-coincidence (A)} A lunar pulse should be visible in only one of the two
beams. An anti-coincidence trigger is set up by excluding events for which a pulse
was found in both beams in the same time trace.

\end{itemize}
Figure \ref{distribution} displays distributions of $S$ after application of
only the timer cut (T), the timer and width cut (TW), and a combination of all
cuts (TWA). The line enclosing the black area in Fig.~\ref{distribution}
corresponds to the distribution of triggers that are expected if the background
is pure Gaussian noise (see Appendix \ref{app:stat} for details of the
calculation). After all cuts have been applied the number of triggers for which
$S>23$ is a factor of 3-4 higher than the amount of triggers expected for
Gaussian noise, while the largest $S$ value in the distribution is about three
times as large as the highest $S$ value for Gaussian noise. Apparently, the
background includes pulsed noise that produces triggers and contains pulses that
are narrow enough to survive the cut on width. The properties of these pulses are
further explored in Sec.\ \ref{sec:peak}.  Due to the pulsed noise, the limit
that we derive for the neutrino flux is less stringent than estimated in
\citet{scholten}, where the existence of this background was not anticipated.

For reference, \figref{twowidths}
shows the difference between the tails of the distribution of $S$ for a $W<10$
and a $W<12$ cut. In the latter case the highest value of  $S$ is larger, but the
corresponding decrease in detection efficiency makes this cut unfavorable (see
\secref{simulations})

\section{Simulations}
\label{sec:simulations}
\subsection{Pulse recovery}

The efficiency with which pulses are found by the analysis procedure and the
effects of data cuts and ionospheric dispersion are simulated by adding pulses to
raw data. The received power and the power after RFI reduction are different for
all bands and polarizations and change with time due to the dynamic gain matching
in the electronics of WSRT. To correct for this the pulses are normalized
following the same procedure as in the analysis described in the previous
sections i.e.\ the pulse strength, denoted by $S_i$, is expressed in
dimensionless units as defined in \eqref{S}.  These pulses are delta peaks
inserted at random times with a random phase. Because the pulses are band
width limited, the bulk of the power in such a pulse typically spreads out over
a few time samples (see Appendix A).  The pulse is dispersed corresponding to a
particular TEC value, named simTEC, and the amplitudes are rounded off towards
nearest integer within the dynamic range, after adding it to the raw
(i.e.~before RFI mitigation) data. For the simulations we have inserted 1000
pulses in a few different 10 seconds segments of raw WSRT data.

\begin{figure}
\centering
\includegraphics[width=0.8\linewidth,viewport=20 35 520 510,clip]{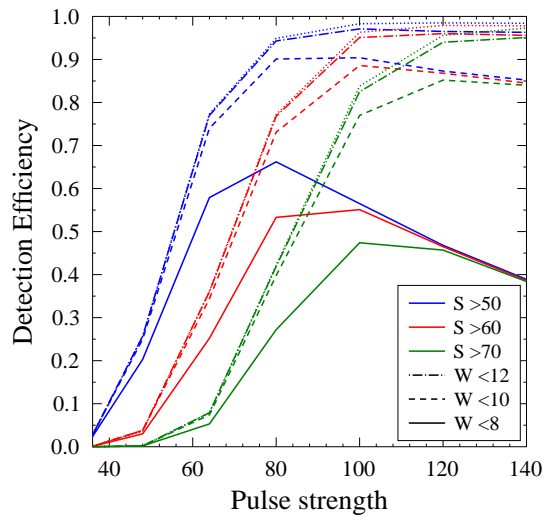}
\caption{The detection efficiency is shown as function of pulse strength for various settings of the trigger conditions as discussed in the text.
The colors correspond to different values for the pulse-strength thresholds $S_{th}$ and the line styles correspond to different maximum widths.
The dotted curve indicates the efficiency when no width cut is applied.
All pulses are simulated with simTEC =12 and analysed with TEC =10.}
\label{fig:x-pol}
\end{figure}

We define the detection efficiency (DE) as the fraction of inserted pulses that
is retrieved after applying the trigger conditions and the cuts that are used in
the analysis. \figref{x-pol} shows the DE for inserted pulses of strength varying
from $S_i=36$ to $S_i=140$. Each pulse is inserted in the x-polarization and
dispersed with a simTEC=12. The de-dispersion is done with STEC=10 to simulate a
practically unavoidable error in the STEC value. The blue, red, and green lines
in \figref{x-pol} show the DE for recovering pulses with strength exceeding
$S_{th} >50$, $S_{th} >60$ and $S_{th} >70 $ respectively. Due to interference
with the background the recovered pule strength differs from the input value
$S_i$. The dotted lines show the DE without any width cut applied. Solid lines
represents the DE with width cut $W< 8$, dashed lines show width cut $W<
10$, whereas width cut $W < 12$ is shown by dash-dotted lines.

From \figref{x-pol} it can be seen that the DE tends to saturate to unity for
large pulses, as is to be expected. However one also sees that the width cut may
severely limit the DE which even worsens with increasing $S_i$. The reason for
this is that with increasing pulse strength the width (as defined in this work)
increases. For really large pulses the signal may saturate causing an additional
broadening of the recovered pulse. In general one also sees that the input pulse
has to be about 10 units in magnitude larger that the threshold to be recovered
with more than 50\% efficiency.

\begin{figure}
\centering
\includegraphics[width=0.8\linewidth,viewport=80 90 620 595,clip]{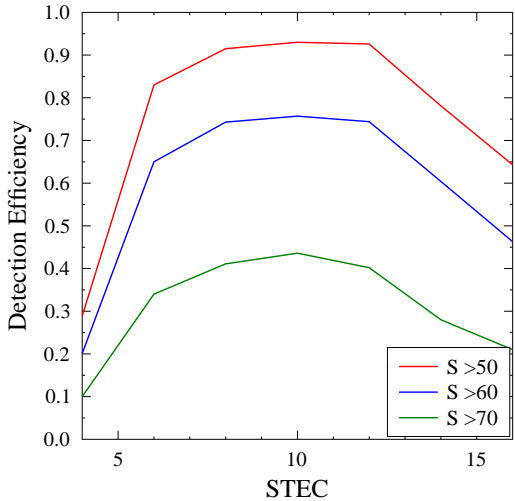}
\caption{The detection efficiency for pulses of strength $S_i=80$ and simTEC=10
is shown as function of the STEC value used in the analysis.
The colors correspond to different thresholds. }
\label{fig:tec-xpol}
\end{figure}

The effect on the DE of the difference between the STEC value used in the
generation of the pulse (simTEC) and the value used in the analysis is studied in
\figref{tec-xpol} for pulses of strength $S_i=80$, width cut $W<12$ and
simTEC=10. There are two effects playing a role here. Firstly, a larger error in
the STEC results in a more dispersed pulse for which the power is divided over
more time samples.
Secondly, the range of time samples scanned in the different frequency bands
after a trigger has been found in the first band, depends on the STEC value. If
simTEC is much smaller than the actual STEC value, the pulse may be located
outside the scanning range and is not found. When simTEC is higher than the
actual value, this problem does not occur, causing an asymmetry in
\figref{tec-xpol}.

\begin{figure}
\centering
\includegraphics[width=0.8\linewidth,viewport=82 120 578 610,clip]{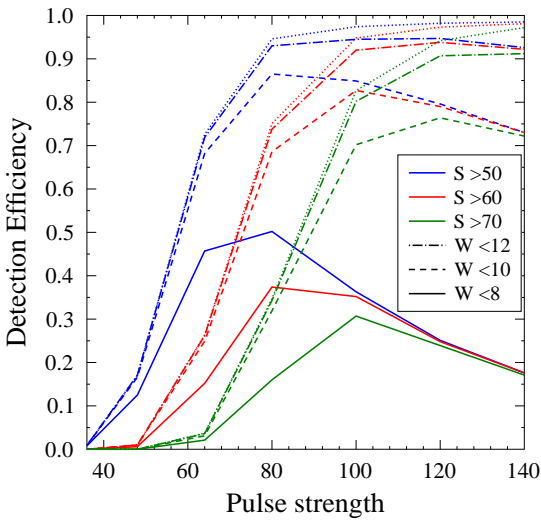}
\caption{Same as \figref{x-pol} however a Faraday rotation of the polarization direction is taken into account.}
\label{fig:faraday}
\end{figure}

Due to the presence of the Earth magnetic field, the polarized radio signal is
subject to a Faraday Rotation in the ionosphere (see \appref{Far}) which induces
a rotation of the angle of linear polarization across the frequency band. On
Earth the pulse will thus be polarized in the x-direction for certain frequencies
while in the y-direction for another frequency. In each polarization direction
the signal will thus cover a rather limited band width causing a broadening of
the pulse and thus  to a possible decrease in the DE. \figref{faraday} shows the
DE as function of the pulse strength including the effect of Faraday rotation. A
rather large decrease in the DE is seen when width cuts are applied. This
behavior is observed for all values of $S_{th}$. We therefore adopt the $W<12$
cut to be used on the analysis of the data.

\begin{figure}
\centering
\includegraphics[width=0.8\linewidth,viewport=80 120 540 560,clip]{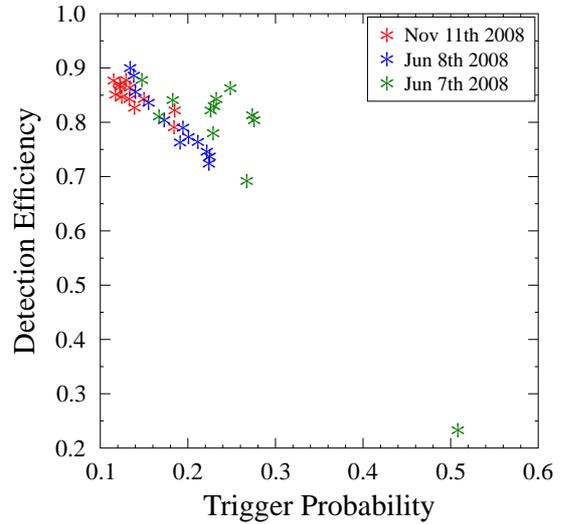}
\caption{The detection efficiency for pulses added with strength 25, simTEC 12,
STEC 10 to different data traces of 10~s each is plotted v.s.\ the number of triggers in the
data trace before addition of the pulse.
The colors correspond to observation runs at different dates.}
\label{fig:anticorrelation}
\end{figure}

We have observed a sizable dependence of the determined DE on the data trace
which was used. First we have investigated a possible correlation of the DE value
with RFI-power but this was not conclusive. We found however a very pronounced
correlation between the DE and the number of raw triggers in the time trace. To
show this we have determined the DE for a wide selection of raw time traces. We
concentrated on the case where pulses of strength $S_i=100$, simTEC=$12$,
STEC=$10$, were added to time traces taken from observations of June
$7^{th}$~2008, June $8^{th}$~2008, and Nov. $11^{th}$~2008. We have processed 12
consecutive traces each of 10~s. The pulses above a threshold of $S_{th}=65$ were
recovered using a width cut $W<12$. We found a clear anti-correlation between the
DE and the raw trigger rate, see \figref{anticorrelation}. On the basis of this
analysis the data of June $7^{th}$~2008 and June $8^{th}$~2008, have been
excluded from analysis because of the large number of triggers. Other
observations show a raw trigger rate of less than 40 per time trace of 0.1~s. The
observed correlation can be understood from the fact that if the algorithm finds
a pulse in a spectrum, this spectrum will not be searched any further for the
occurrence of another pulse. Thus if each spectrum in a 0.1~s time trace
generates a trigger, the chance of recovering the added pulse will be vanishingly
small.

\section{Effective observation time}

\label{sec:background} The effective observation time is decreased by a number of
effects. After we have excluded the two observation runs which harbour
exceptionally large numbers of raw triggers, we have 51.1 hours of dual-beam observation time left.
In the data analysis some files are missing or are not usable due to various
reasons, such as hardware and software failures, corresponding to a loss of 3.5
hours.

When a trigger is found the rest of the time trace is not scanned for pulses. Per
raw trigger this corresponds to a mean lost time of 250 $\mu$s (single beam).
Because of the coincidence cut, the whole time trace in the other beam should
also be counted as lost time, resulting in another 500 $\mu$s (single beam). For
7.6 million raw triggers this adds up to 0.8 hours (dual beam).

The cut on the timer pulses should be regarded as cutting out observation
time, but this has already been accounted for in the previous step. Each time
the system triggers on the timing pulse 750 $\mu$s (single beam) is lost, as is
the case for any other trigger.

After RFI removal the first and last 250 samples of a time trace have to be neglected due to FFT edge effects, corresponding to a 0.25\% loss
of observation time.

The total observation time is therefore $(51.1-3.5-0.8)\times 0.9975= 46.7$ hours
of dual beam data. Each beam covers about a third of the lunar area.

\section{Background}

For a radio antenna, the 1 $\sigma$ noise power density $F_n$ [Jy] is given by
\begin{equation}
F_n = { 2 k T_{\mathrm{sys}} \over \sqrt{\Delta t \Delta \nu} A_{\mathrm{eff}}} 10^{26} \mathrm{Jy},
\end{equation}
where $k$ is the Boltzmann constant ($1.38 \times 10^{-23}$ Joules), $T_{\mathrm{sys}}$ is
the antenna effective temperature in Kelvins, $\Delta t$ and $\Delta \nu$ are the
time and frequency bins of the measurement, and $A_{\mathrm{eff}}$ is the effective area of the telescope in m$^2$. For the 14
Westerbork antennas we used, $A_{\mathrm{eff}}=491$~m$^2$. For the measuring band which we need, the LFFE
measuring between 113 and 170 MHz, one has a $T_{\mathrm{sys}}$ of 400 - 700 K. We use 11
antennas, yielding for the noise power per Nyquist time sample, $\Delta t \Delta
\nu=0.5$,
\begin{equation}
 F_n = { 2 \times 1.38 \times 10^{-23} \over 11\sqrt{0.5}}
 10^{26} \left(\frac{T_{\mathrm{sys}}}{491}\right)= 349 \left(\frac{T_{\mathrm{sys}}}{491}\right) \mathrm{Jy},
\end{equation}
which covers the range 286 - 500~Jy. In the following we have adopted the value
of $\sigma^2=400$~Jy as the average of the observing bandwidth.

\section{Results}\seclab{results}

\begin{figure}
\centering
\includegraphics[width=0.7\linewidth,viewport=83 120 570 595,clip]{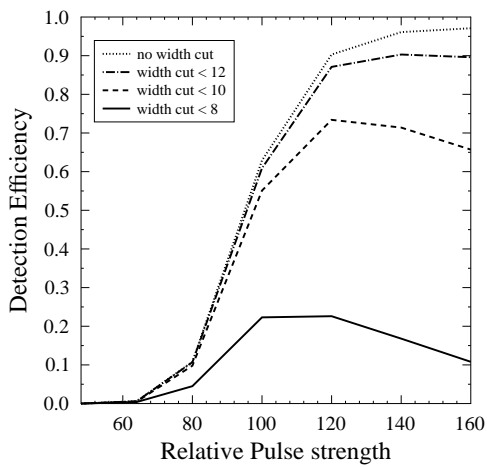}
\caption{ The detection efficiency for a detection threshold $S_{th}=77$ is shown as a function of pulse
strength for various settings of the trigger conditions as discussed in the text.
The dotted curve indicates the efficiency when no width cut is applied. The effect
of Faraday rotation has been taken into account
}
\label{fig:farad}
\end{figure}

In 46.7 hours of data no triggers were found with a strength exceeding
$S_{th}=77$. To convert this into a probability for not observing the Moon we
calculate the DE curve for a detection threshold $S_{th}=77$ including the
effects of Faraday rotation, see \figref{farad}. From this figure it can be seen
that the DE reaches a value of 87.5\% for pulses in excess of $S_i > 120$ and
width cut $W< 12$. This corresponds to 120$\sigma^2\times 5 = 240$~kJy. For
pulses of lower strength the DE drops rapidly due to interference with the
background. For comparison we will also consider the case for detecting pulses
with a strength of $S_i > 90$ for which the DE has dropped to 50\%.

The lack of pulses stronger than a certain magnitude implies a new limit on the
flux of ultra-high energy neutrinos. To obtain the limit requires a calculation
of the acceptance which takes into account the attenuation of the radio signal
inside the Moon, the transmission at the lunar surface and the angle with respect
to the arrival direction of the neutrino. On basis of the simulations which are
described in \citet{scholten}, the 90\% confidence level flux limit has been
determined. In arriving at this the model-independent procedure described in
\citet{forte} has been followed.

\begin{figure}
\centering
\includegraphics[width=0.95\linewidth,viewport=50 150 550 700,clip]{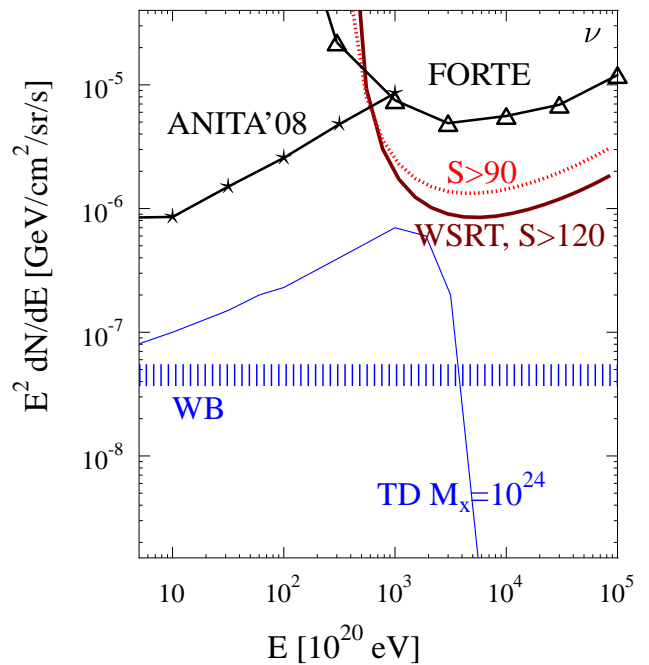}
\caption{Neutrino flux limit currently established with 46.7 hours of WSRT data.
The brown (red) line is calculated for a minimum pulse strength of $S=120$ ($S=90$),
corresponding to a DE of 87.5\% (50\%).
Limits set by ANITA~\citep{anita08} and FORTE~\citep{forte}
are included in the plot as well as the Waxman-Bahcall flux~\citep{Waxman:1998yy} and a TD model prediction~\citep{ps96}.}
\figlab{wsrtlimits}
\end{figure}

In \figref{wsrtlimits} the 90\% confidence limits are given for the two cases we
have analyzed, $S_i > 120$ corresponding to a DE=87.5\% and $S_i > 90$ (DE=50\%).
As can be seen the gain in DE is far more important in setting the neutrino flux
limit than the loss in sensitivity. Only at the lowest neutrino energies the
result is reversed. In the rest of this work we will therefore base all arguments
on the $S_i > 120$ limit. In arriving at this limit the same assumptions have
been made as in \citet{scholten}, in particular that the neutrino cross sections
equal the prediction given in \citet{Gan00}.

The current limits in the UHE region are established by ANITA \citep{anita08} and
FORTE \citep{forte}. Near the bottom of  \figref{wsrtlimits} two model
predictions are plotted, the Waxman-Bahcall limit \citep{wb01} and a top-down
model \citep{ps96} for exotic particles of mass $M_X=10^{24}$~eV.

Calculation of the flux of UHECRs and UHE neutrinos from the decay of topological
defects is very model dependent. Parameters of such scenarios include, mass of
the topological defect $M_X$, energy spectra, and final state composition of the
decay products, and cosmological evolution of the injection rate of topological
defects. The freedom provided by the reasonable range of values of these
parameters is constraint by limits on the gamma ray flux at GeV-TeV energies and
neutrinos at energies above $10^{20-21}$~eV. The curve plotted in
\figref{wsrtlimits} corresponds to a $M_X=10^{24}$ scenario based on \citet{ps96}. If future limits can constrain the
neutrino flux by another order of magnitude this will put constraints on the
degrees of freedom of top-down models.

\begin{figure}
\centering
\includegraphics[width=0.95\linewidth,viewport=50 150 550 700,clip]{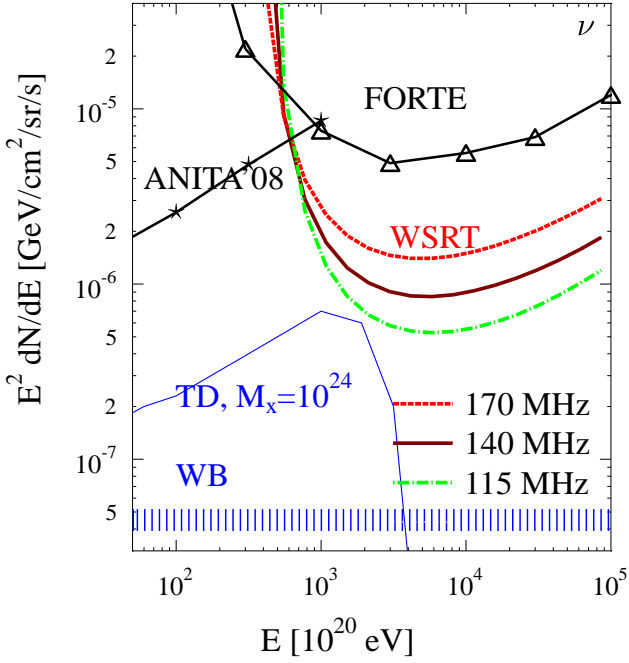}
\caption{Same as \figref{wsrtlimits}. The limits are calculated for the central frequency and
the upper and lower limit of our band-width.
}
\figlab{wsrtlimits-nu}
\end{figure}

The acceptance calculations have been done at a frequency of 140\,MHz which is
central in the observing bandwidth. Since the acceptance depends on the third
power of the frequency, it varies considerably over the bandwidth as shown in
\figref{wsrtlimits-nu}, however the average agrees with the calculation at
140\,MHz.

The systematic error on the acceptance is dominated by three uncertainties: the
density profile, attenuation length, and stopping power of the lunar regolith.
Some of these errors have been estimated in Ref.~\cite{scholten}. In particular
the effect of density was considered which is rather complicated as an increase
in the density reflects in an increased value for the index of refraction, a
shorter shower length and thus a larger angular spread, more attenuation, and a
lower mean neutrino interaction depth. Many of these effects appear to compensate
each other resulting in an acceptance that is almost density independent, only
slightly raising the minimal energy for neutrino detection. The error due to
unknown variations in the density profile is therefore estimated to be 10\% in
threshold energy (which is the difference between the 'full' and the 'rock' calculation in Fig.~10 of \citet{scholten}).

In \citet{os75} the loss tangent is expressed in terms of the FeO and TiO$_2$
content of the samples. On the basis of this we arrive an uncertainty in the
attenuation length for radio waves of about 40\% which directly reflects in a
similar error in the flux determination. Also the stopping power of the regolith
depends on the chemical composition
where we have used a radiation length of 22.1\,g/cm$^2$. An typical variation of the radiation length for the lunar regolith amounts to 0.5\,g/cm$^2$. Since the angular spread is proportional to the shower length and the acceptance goes with the third power of the spread, this corresponds to a variation in the acceptance of 10\%.
As argued before, surface roughness is not very important at our wavelength and may contribute not more than 10\% to the uncertainty in determining the flux.
The error in the Moon coverage
of the two beams is estimated at 20\%. Adding these errors in quadrature gives a
systematic error on the flux of 50\% as indicated in \figref{wsrtlimits-err}.

\begin{figure}
\centering
\includegraphics[width=0.95\linewidth,viewport=50 150 550 700,clip]{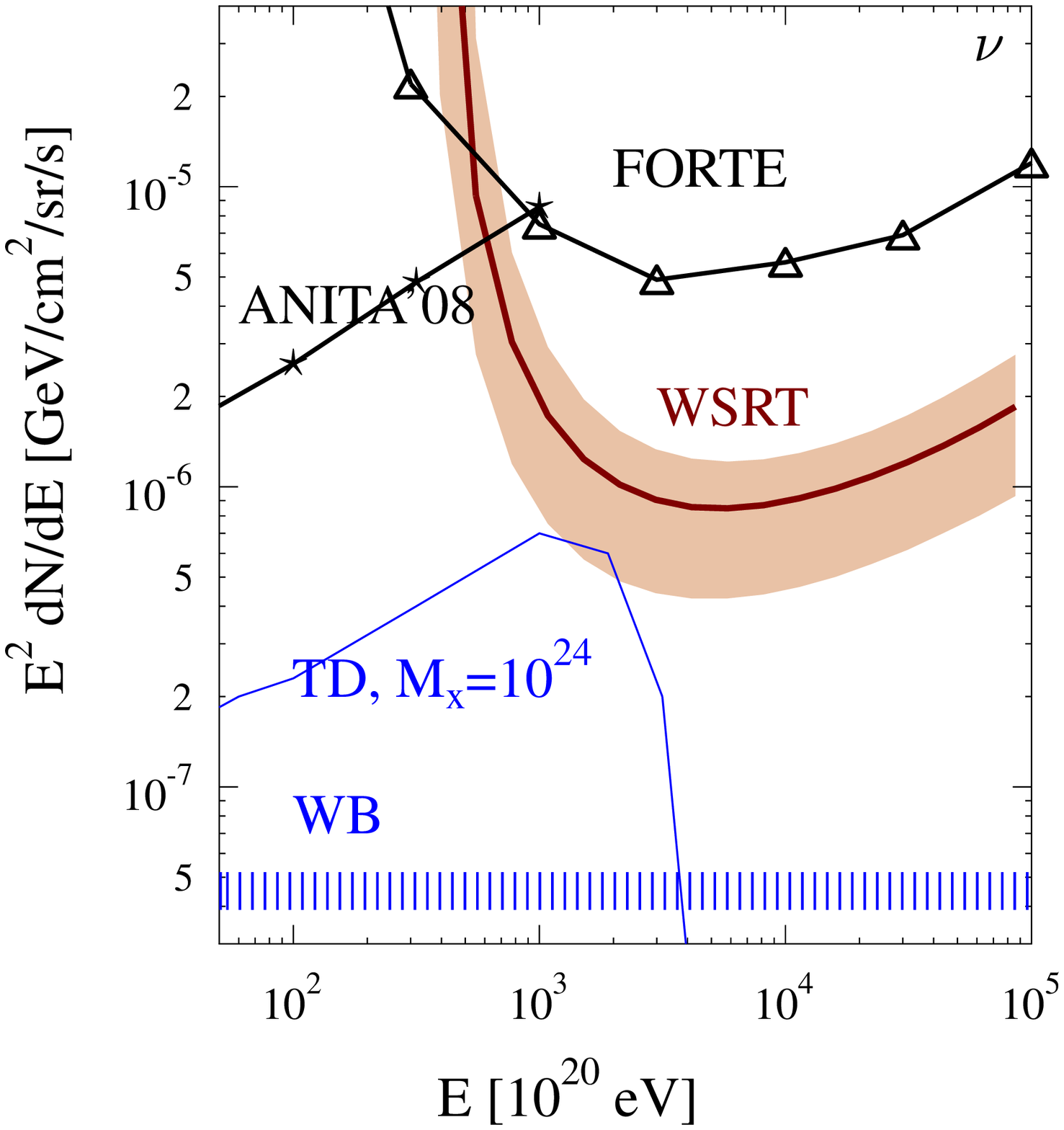}
\caption{[color online] Same as \figref{wsrtlimits}.
The band shows the systematic error.}
\figlab{wsrtlimits-err}
\end{figure}

From the non-observation of short radio pulses coming from the Moon, limits can
also be set on the flux of UHE cosmic rays. This will be discussed in a future
article as special attention has to be devoted to the calculation of the
formation length for Cherenkov radiation which is important for a shower that is
close to the lunar surface.

\section{Discussion of the large peaks}
\label{sec:peak}
\begin{figure}
\centering
\includegraphics[width=0.45\linewidth]{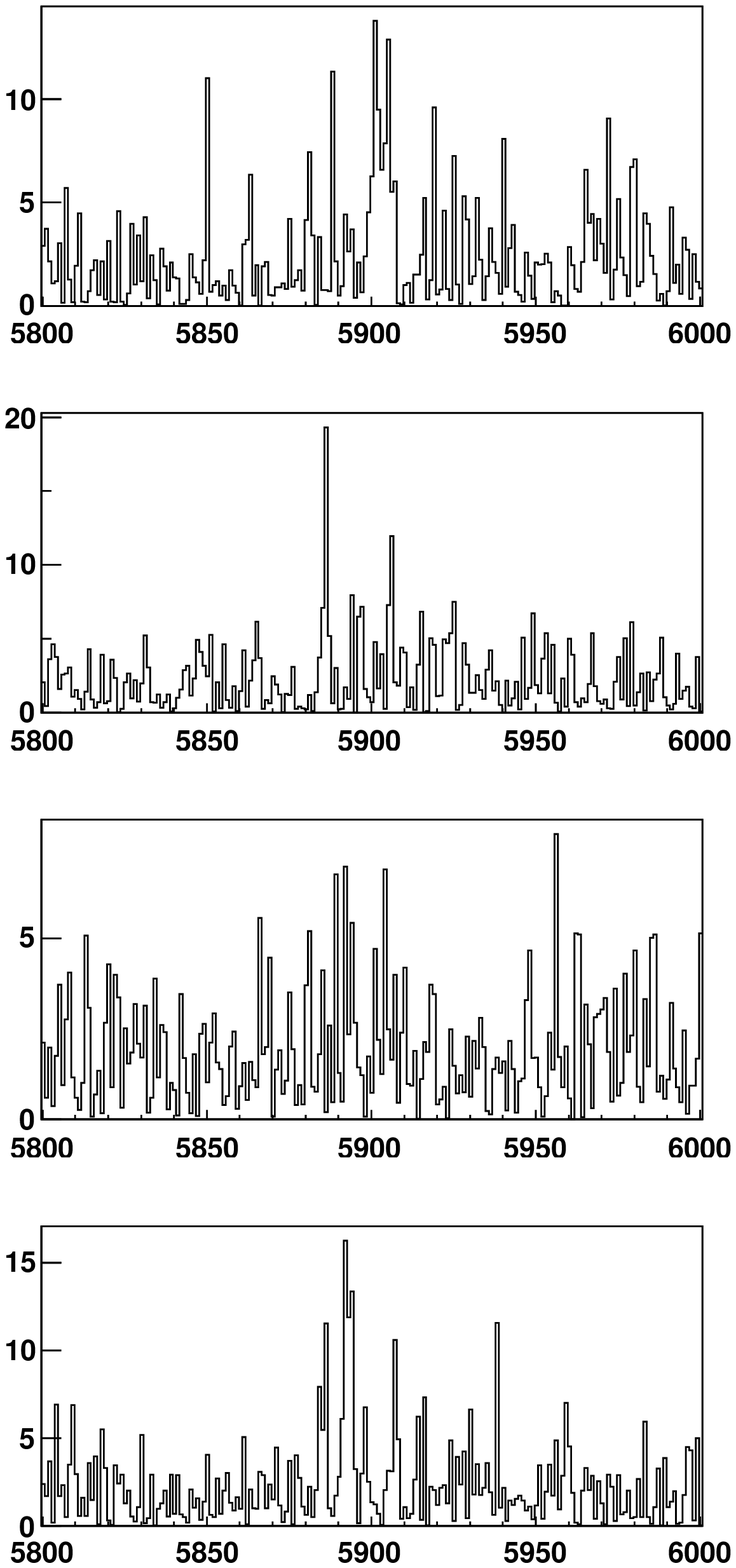}
\includegraphics[width=0.45\linewidth]{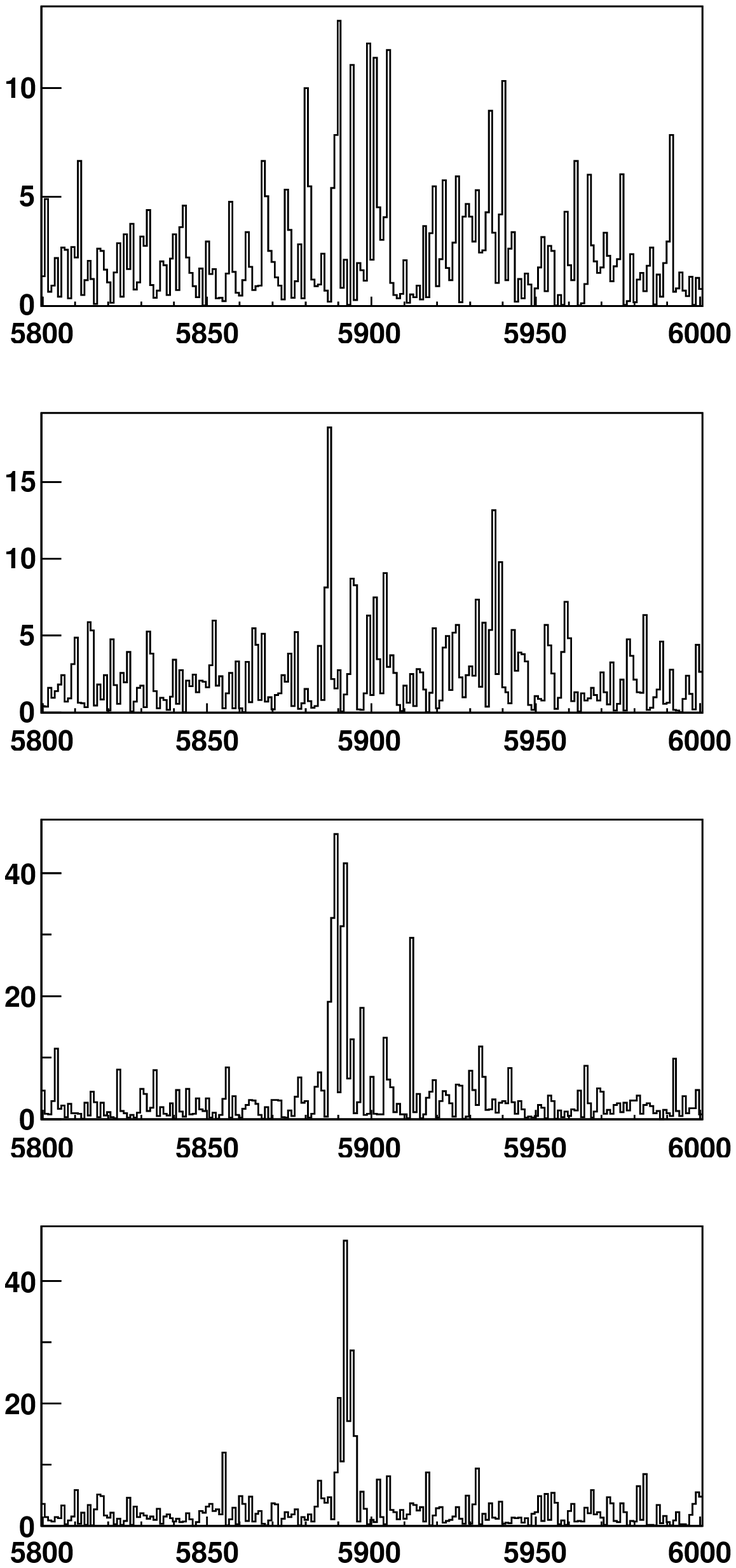}
\caption{Time traces of the power (polarizations added) for a typical large pulse seen in the data after applying the cuts. The P5 values are plotted for all bands (top: highest frequency, bottom: lowest frequency) and both beams (left and right). The horizontal axis displays bin number (bin size is 25 ns). The power on the vertical axis is expressed in mean P5 value, with the trigger level at 5 for all bands. A trigger was only found for the right beam.}
\label{bigpulse1}
\end{figure}

\begin{figure}
\centering
\includegraphics[width=0.45\linewidth]{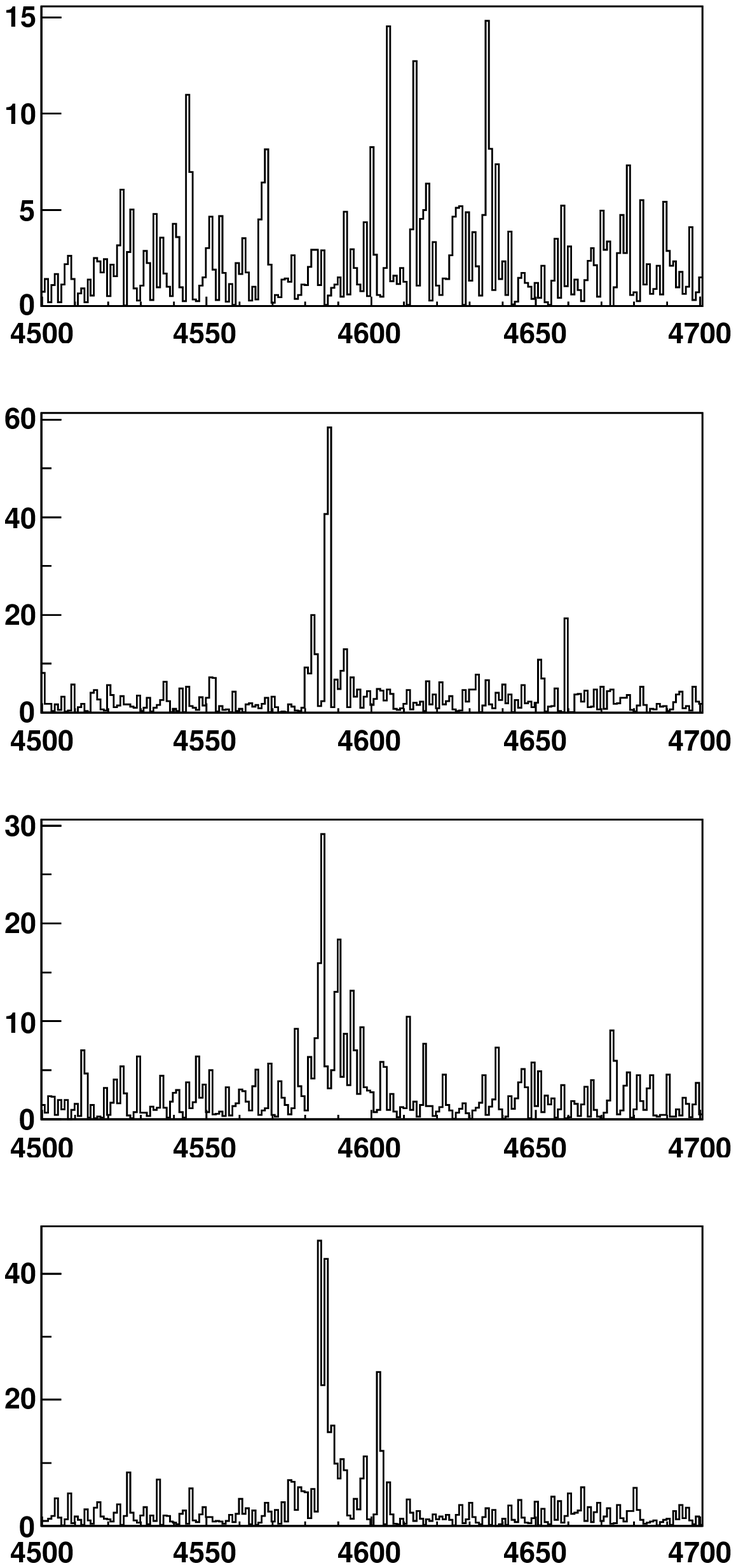}
\includegraphics[width=0.45\linewidth]{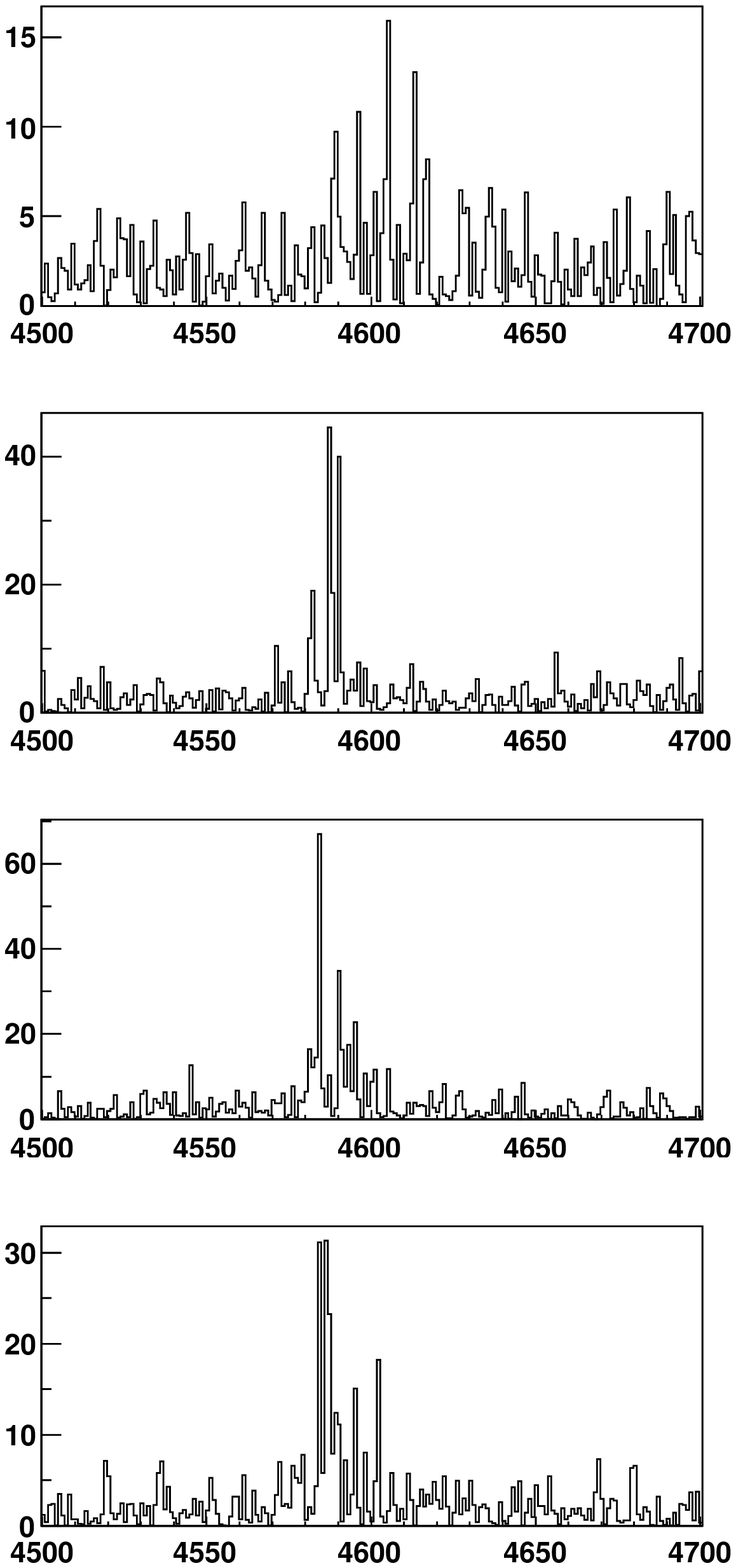}
\caption{Like Fig.~\ref{bigpulse1}, different pulse. The similar features in both beams exclude the pulse as a proper lunar pulse candidate.}
\label{bigpulse2}
\end{figure}

We have investigated the nine strongest pulses that survive the applied cuts.
Figures \ref{bigpulse1} and \ref{bigpulse2} are typical examples of the time traces of such
pulses. The pulses are from different observation runs and their $P5$ values are plotted as a function of bin number (bin size is 25 ns) for all frequency bands and both beams.
At this stage the RFI has already been mitigated and the signal has been de-dispersed.

For both example events, the trigger was found in the right-hand beam. For the event in Figure \ref{bigpulse1} the maximum $P5$ value increases with decreasing frequencies. This could be due to a stronger signal at lower frequencies or an increase of pulsed background at higher frequencies (remember that the $P5$ value is normalized over a 500~$\mu$s time trace for each individual band). Although the signal is clearly much smaller in the left beam, it should be noted that three out of four bands actually have a pulse that exceeds trigger level ($P5 > 5$). The event displayed in Fig.~\ref{bigpulse2} has a strong signal in both beams and the only reason this event was not discarded by the anti-coincidence criterion is that the highest frequency band has a very small signal-to-noise ratio. In this band, the signal is suppressed and happens to be just above threshold in the right-hand beam but below threshold in the left-hand beam. This way, strong temporary increases in background radiation are responsible for several of the largest events that pass our criteria.

Although the event in Figure \ref{bigpulse1} has a curious dependence on frequency, it has the properties of a proper
lunar pulse of the type we are looking for, which are: i) present in all frequency
bands, ii) strong polarization, iii) short after dispersion correction, and iv)
present in one beam only. In order to study the possibility of the pulse to originate from the Moon we can 
impose an additional condition that the Faraday rotation of the polarization is
of the correct magnitude.

The Faraday rotation going through a
plasma with STEC=5 in the Earth magnetic field is about $\pi/4$ radian over 30
MHz (corresponding to the difference in the centroids of bands 1 and 3) at the
frequencies of interest for the present study.
For a pulse fully polarized in the
x-direction in the center of band 1 one would thus expect about equal strength in
the x and y polarization for band 3.
We have examined whether or not the ratio between the pulse strengths in the x and y polarization in the different frequency bands corresponds to the ratios expected on the basis of the STEC value. 
This criterion disqualifies the pulse in Figure \ref{bigpulse1} as originating from outside the ionosphere.

We have studied the nine strongest pulses with $S>62$ and found that all of them are unlikely to come from the Moon, because they either have a strong signal in both beams or do not have the frequency dependent behavior expected from Faraday rotation. As a result we can safely state that we see no pulses originating
from a particle cascade in the Moon with a strength larger than $S=62$. Because this analysis is done a posteriori, the threshold used for the determination of the neutrino flux limit 
is kept at $S=77$. In future studies, additional cut criteria based on temporary power surges in the background and Faraday rotation of the signal in the ionosphere can be implemented to further understand and reduce the background.

\section{Outlook}
The next phase in the NuMoon experiment will be to use LOFAR, the Low Frequency Array \citep{lofar}, that is under construction in
the Netherlands. LOFAR is a network of low frequency omni-directional radio antennas communicating over a fiber optics
network. It will feature two types of antennas operating at different
frequencies,
the Low Band (LB) antennas cover a band of 30--80~MHz while the High Band (HB) antennas cover
the regime 110--240~MHz. The latter will be used for the NuMoon observations.
LOFAR is organized in 35 stations each containing 48 LB and 96 HB antennas. Half of the
stations are located inside the 2~km$\times$2~km core with a total collecting area of $\sim$0.05~km$^2$.
Multiple beams can be formed to cover the surface of the Moon, resulting in a
sensitivity that is about 25 times better than the WSRT \citep{singh08}.


\bibliographystyle{aa}
\bibliography{14104}

\begin{appendix}
\section{Bandwidth limited pulses}
\label{app:pulses}
To give an idea of the behaviour of bandwidth limited, Nyquist sampled pulses we show the sampling of pulse with the shape of a delta peak.
Like in our analysis we take a time trace of 20\ 000 timebins, where each timebin is 25 ns wide. We add a pulse at a random location with a
random phase by setting the amplitude of all frequency bins to the same positive value and the phase $\Phi_i$ of the
frequency bin $i$ to
\begin{equation}
\Phi_i = \Phi_0 + 2\pi i \left[ 1-\left(\frac{t}{20000}\right)\right]
\end{equation}
where $\Phi_0$ is a random phase and $t$ is a random time between 0 and 20\ 000. The shape of the pulse in time domain is found by performing a reverse FFT.
For $\Phi_0=0$ and an integer value of $t$, all timebins are zero except timebin $t$, which contains a positive value. For a random $\Phi_0$ or a non-integer value of $t$ the
pulse has a complicated shape in the time domain.

\begin{figure}
\centering
\includegraphics[width=7cm]{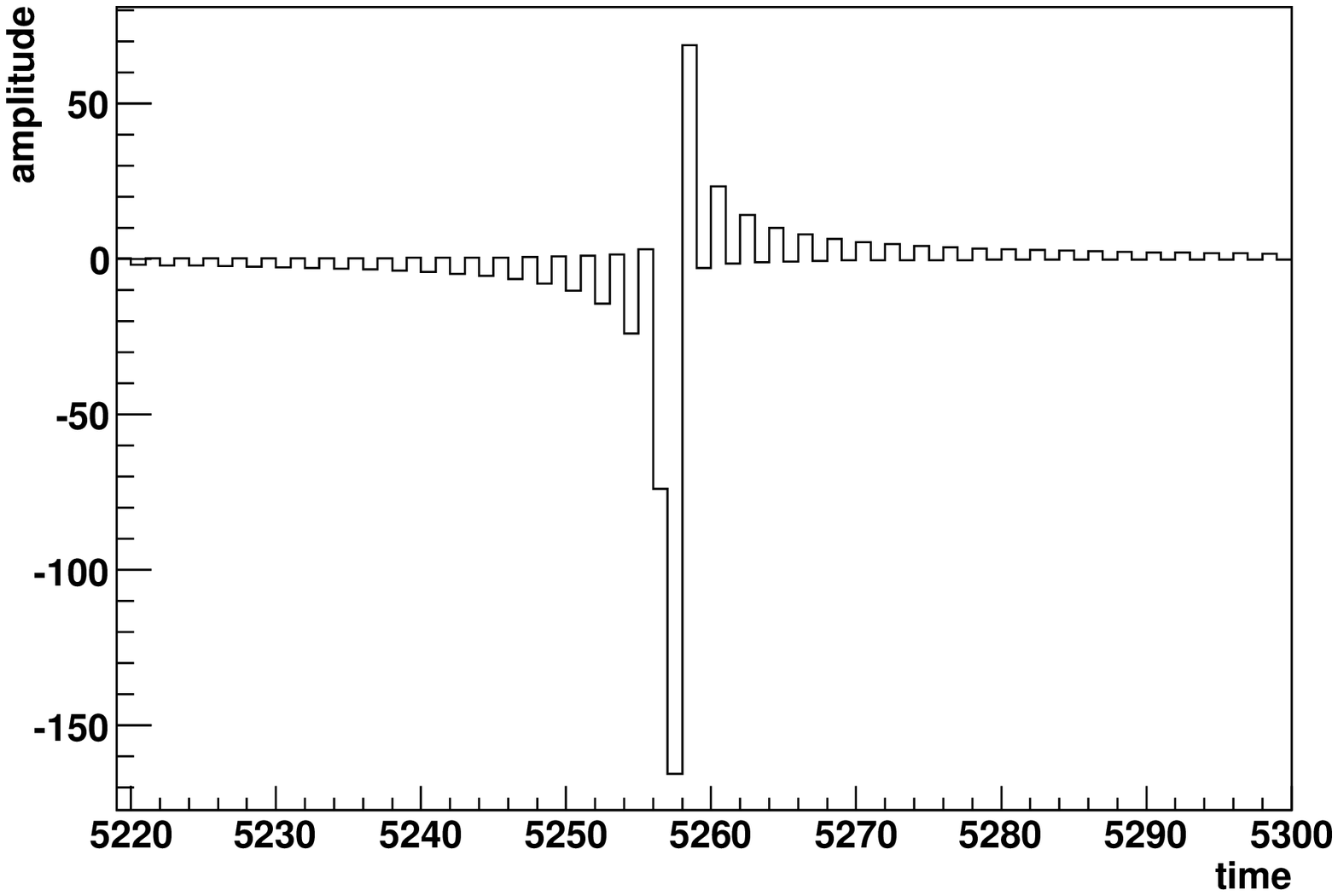}
\includegraphics[width=7cm]{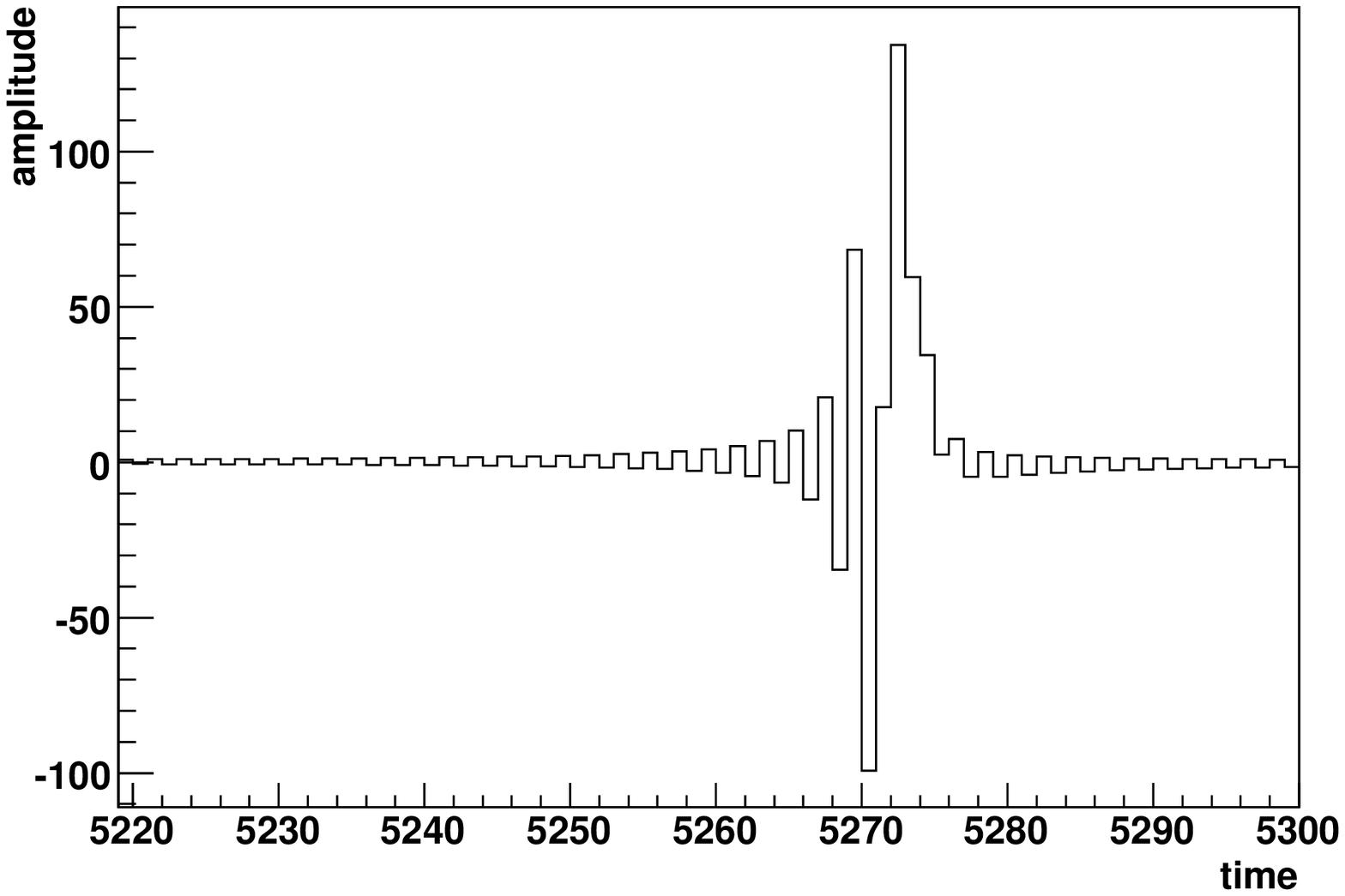}
\includegraphics[width=7cm]{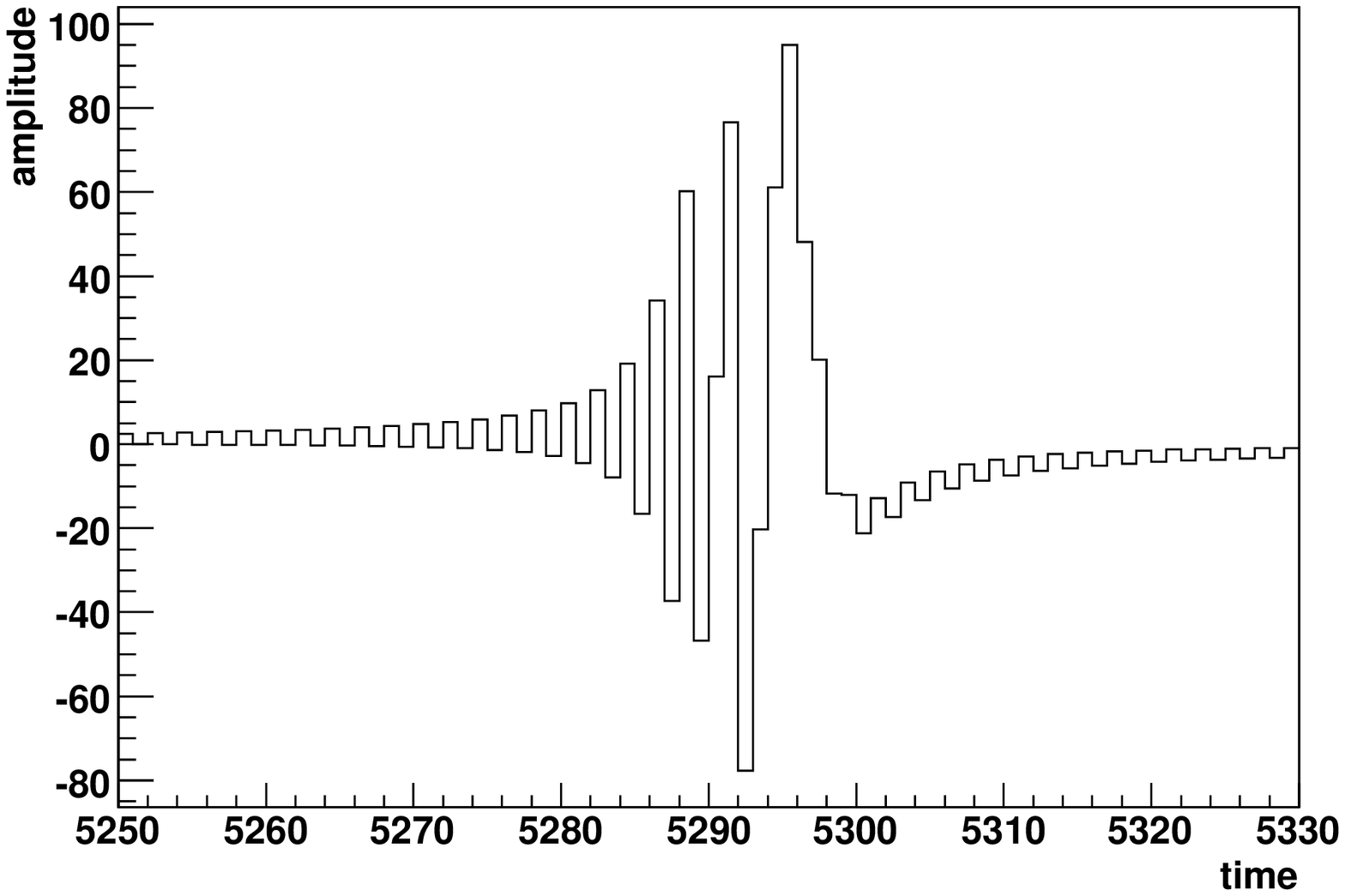}
\caption{Top: Time sampling of a delta peak pulse at time 5257.96 with phase 1.17$\pi$.
Middle: same pulse, dispersed with TEC value 4. Bottom:
Same pulse dispersed with TEC value 10.}
\label{pulses}
\end{figure}

The top panel of Figure \ref{pulses} shows how such a pulse typically spreads out over many timebins. In most cases the bulk of the pulse power is
inside 2 or 3 timebins. When the pulse is dispersed the power is spread out over even more timebins. The middle and bottom panel show the pulse
broadening for TEC values of 4 resp.\ 10. In our analysis we have defined the width of the pulse as the number of consecutive $P_5$ values that
exceed the threshold. It should be noted that this differs in general from the actual width of the pulse. For example, a very large amplitude in
one single timebin can give 5 consecutive $P_5$ triggers, while the same pulse spread out over 5 timebins will maybe produce only one threshold
exceeding $P_5$ value.

\section{Statistics}
\label{app:stat}
We carry out a statistical analysis to establish the expected amount of triggers and the distribution of $S$, the sum of the $P_5$ values
over 4 frequency bands, for
Gaussian background noise. When the amplitudes follow a Gaussian distribution, the probability density of the power is a
chi-squared distribution
\begin{equation}
P(x,k)=\frac{x^{k/2-1}e^{-x/2}}{2^{k/2}\Gamma(k/2)},
\label{probdens}
\end{equation}
where $x$ is the power in units of the standard deviation $\sigma$ and $k$ is the
number of degrees of freedom. The trigger condition for a single band is $P_5>5$, where $P_5$ is given by Eq.~\ref{P5}. For Gaussian
noise with the same standard deviation $\sigma$ in both polarizations this condition becomes
\begin{equation}
P_5=\frac{\displaystyle\sum_{\mathrm{10\ bins}} x}{\bigg<\displaystyle\sum_{\mathrm{5\ bins}} x\bigg>}=
\frac{1}{5} \displaystyle\sum_{\mathrm{10\ bins}} x >5,
\label{p5crit}
\end{equation}
or
\begin{equation}
\displaystyle\sum_{\mathrm{10\ bins}} x>25.
\end{equation}
The trigger chance for a single frequency band is therefore
\begin{equation}
P_{\mathrm{trigger}} = \int^{\infty}_{25} P(x,k=10)\, \mathrm{d}x \approx 0.00535.
\end{equation}
Consecutive $P_5$ values have 8 overlapping timebins (4 in both polarizations) and we have to distinguish the chance to find a
trigger that comes after another trigger, $P_{\mathrm{xx}}$, and a trigger after a non-trigger, $P_{\mathrm{ox}}$.
A trigger will be found after a non-trigger if:
\begin{itemize}
\item The eight overlapping bins add up to less than the trigger value by a certain value $A$.
\item The two bins for the non-trigger add up to a value \emph{smaller than $A$}.
\item The two bins for the trigger add up to a value \emph{larger than $A$}.
\end{itemize}
To find the total probability we integrate over all possible values of $A$
\begin{eqnarray}
P_{\mathrm{ox}} & = & \nonumber
\int_{0}^{25} P(25-A,k=8) \left[\int_{0}^{A} P(x,k=2)\, \mathrm{d}x \right] \\ & & \left[ \int_{A}^{\infty} P(x,k=2)\, \mathrm{d}x
\right] \mathrm{d}A \\
&\approx &0.00283 \nonumber  \approx 0.53 P_{\mathrm{trigger}} \nonumber.
\end{eqnarray}
The chance of finding a trigger after another trigger is
\begin{equation}
P_{\mathrm{xx}} = P_{\mathrm{trigger}} -P_{\mathrm{ox}} \approx 0.00252 \approx 0.47 P_{\mathrm{trigger}},
\end{equation}
so about half of the triggers is found clustered together which has consequences for our analysis.

For a complete trigger the trigger condition has to be met in all 4 frequency bands. In bands 2, 3, and 4 a range of $P_5$
values will be scanned based on the STEC value. Suppose in the first band a trigger is found after a non-trigger. The chance
to find a trigger in band no.\ $i$ is
\begin{equation}
P_i = P_{\mathrm{trigger}} + (N_i -1) \cdot P_{\mathrm{ox}} + \mathcal{O}(P_{\mathrm{\mathrm{trigger}}}^2),
\end{equation}
where $N_i$ is the number of values scanned in band $i$.
Terms of the order of $P_{\mathrm{trigger}}^2$ arise from properly adding the
chances of finding a trigger in one of the $N_i$ bins and deviations in the
chance of finding a trigger in a certain bin depending on the number of bins
without a trigger than precede it. The chance to find a trigger in the three
upper bands is $P_2\cdot P_3 \cdot P_4$ for this case.

When, however, the trigger in band 1 came after another trigger in band 1, the chance to find a trigger in the three
upper bands is
smaller because we know that for the previous trigger in band 1, not all upper bands had a trigger. If that were the case the
pulse search would have skipped the rest of the time trace. The range of timebins that is scanned in the upper bands is
overlaps with scan after the previous trigger in band 1. Actually, in each band only one new timebin is scanned.
The chance to find a
trigger in the upper bands is therefore now reduced to $P_2\cdot P_3 \cdot P_4 \cdot (1-P_{\mathrm{reduce}})$, where $P_{\mathrm{reduce}}$ is the
chance that there was also a trigger in the previous scan. To not have a trigger in the previous scan, at least one of bands
should have the trigger in the last timebin, because this is the only bin that is unique for the new scan. In other words,
the previous scan also had a trigger when for each
upper band the trigger is located in any but the last scanned timebin.  The chance for the trigger to be in the last timebin of band $i$
is $P_{\mathrm{ox}}/P_i$, so we find
\begin{equation}
P_{\mathrm{reduce}}=\left( 1 -\frac{P_{\mathrm{ox}}}{P_2}\right)\left( 1 -\frac{P_{\mathrm{ox}}}{P_3}\right)\left( 1
-\frac{P_{\mathrm{ox}}}{P_3}\right)
+\mathcal{O}(P_{\mathrm{trigger}}),
\end{equation}
where we neglect terms of order $P_{\mathrm{trigger}}$ that arise from the possibility that a trigger is in the last timebin,
but also in the first timebin of the previous scan. The complete chance of finding a trigger in all four bands is given by
\begin{equation}
P_{\mathrm{4trig}}= P_{\mathrm{ox}} P_2 P_3 P_4 + P_{\mathrm{xx}} P_2 P_3 P_4 (1-P_{\mathrm{reduce}})+\mathcal{O}(P_{\mathrm{trigger}}^5) .
\label{notriggers}
\end{equation}
In Table \ref{tab:stat} the values of $P_{\mathrm{4trig}}$ are given for
different STEC values. In the calculation higher order terms are incorporated.
The rightmost column shows the simulated trigger chance that is found by applying
our data analysis code on generated Gaussian noise. In the code the Gaussian
noise is rounded off to the nearest of the 34 values that are part of the dynamic
range of the PuMa-II system. This is the main reason for the discrepancy between
analytic predictions and the simulation results, which is smaller than 5\%.
\begin{table}
\caption[]{Predicted and simulated trigger chance}
\label{tab:stat}

\begin{tabular}{llll}
\hline
STEC & ($N_2$,$N_3$,$N_4$) & pred. chance & sim. chance \\
\hline
5 & (3,5,7) & $1.5\cdot 10^{-8}$ & $1.5\cdot 10^{-8}$ \\
10 & (5,7,11) & $3.6\cdot 10^{-8}$ & $3.8\cdot 10^{-8}$ \\
15 & (5,9,15) & $6.3\cdot 10^{-8}$ & $6.0\cdot 10^{-8}$ \\
\hline
\end{tabular}

\end{table}

To arrive at the (not normalized) probability distribution of $S$, we
take 4 probability distributions of values $a$ through $d$,and integrate to find the distribution of
$x=a+b+c+d$
\begin{eqnarray}
P(x)=
\int_{25}^{x-75} \mathrm{d}a
\int_{25}^{x-a-50} \mathrm{d}b
\int_{25}^{x-a-b-25} \mathrm{d}c\ P(c,k=10)\nonumber \\
P(b,k=10)\ P(a,k=10)\ P(x-a-b-c,k=10) ,
\end{eqnarray}
where the limits of the integral are chosen in such a way that $a$,$b$,$c$ and $d$
all exceed 25 individually. By normalizing the distribution $P(x)$ with the total
amount of triggers projected with Eq.\ \ref{notriggers} and substitute $S=x/5$
(see Eq.\ \ref{p5crit}), we arrive at the analytical noise background prediction
that is plotted in Figure \ref{distribution}.

\section{Ionospheric effects}

The ionosphere is a plasma where the density of free electrons affects the
propagation of electromagnetic waves which may show as a dispersion of the signal
or a frequency-dependent rotation of the linear polarization.

\subsection{Dispersion}

As the radio signal propagates through the Earth's ionosphere it is dispersed by
\begin{equation}
\phi(\nu) = 2 \pi \int \mathrm{d}z \ \nu \left(\sqrt{1-\frac{\nu_{p}^2}{\nu^2}}-1\right) /c,
\end{equation}
where $\Delta \phi$ is the phase shift at frequency $\nu$. The integral is taken
over the traversed distance $z$, $c$ is the speed of light in vacuum, and
$\nu_{p}$ is the the plasma frequency
\begin{equation}
\nu_p^2 = \frac{n_e e^2}{4 \pi^2 \epsilon_0 m_e}=8.07\cdot 10^{17} \mathrm{STEC}/\Delta z,
\end{equation}
where $n_e$ is the electron number density, $e$ the elementary charge,
$\epsilon_0$ the permittivity of vacuum, and $m_e$ the electron mass. For the
ionosphere $\nu_p\approx 3 MHz$ and
\begin{equation}
\int{ \nu_p^2\over 2c } \mathrm{d}z = 1.34 \times 10^{9}STEC,
\end{equation}
where we use the Slanted Total Electron Content (STEC), which is the electron
density integrated along the distance the pulse has traveled through the
ionosphere. The STEC is given in TEC units (TECU) where $1 \mathrm{TECU} =
10^{16}\ \mathrm{electrons}/\mathrm{m}^2$. The phase shift is approximately
\begin{equation}
\phi(\nu) \approx 2 \pi \frac{1.34\cdot 10^{9} \mathrm{STEC}}{\nu},
\end{equation}
corresponding to a time offset of
\begin{equation}
\Delta t = 1.34 \cdot 10^{9} \cdot \mathrm{STEC} \left( \frac{1}{\nu_{1}^2}-\frac{1}{\nu_{2}^2}\right) ,
\end{equation}
between two frequency components $\nu_1$ and $\nu_2$. For an interval of 20 MHz
(140-160 MHz) and STEC=10 the difference in time delay is $\Delta t \approx
1.6\times 10^{-7}$~s, which corresponds to 6.4 time samples for a 40 MHz sampling
frequency.

\subsection{Faraday rotation}\seclab{Far}

In the presence of a magnetic field the linear polarization direction of an
electromagnetic signal will rotate over a finite angle. This Faraday rotation is
usually expressed in terms of a rotation measure (RM)
\beq
\beta_F=RM \lambda^2=RM c^2/\nu^2 \; .
\eeq
In units of radians per square meter (rad/m$^2$), RM is calculated as
\bea
\mathrm{RM} &=& \frac{e^3}{8\pi^2 \varepsilon_0 m^2c^3}\int_0^d n_e B
\;\mathrm{d}s
 = 2.62 \times 10^{-13} \int_0^d n_e B \;\mathrm{d}s \nonumber \\
 &=& 2.62 \times 10^{3}\times STEC
\times B_\parallel
\eea
with B in teslas (T), and $n_e$ in m$^{- 3}$.

The difference in the Faraday rotation angle for two frequency components $\nu_1$
and $\nu_2$ can be related to the difference in time delay as $\Delta\beta_F={e B
\over m_e c}c \Delta t = 5.27 \times 10^6 \Delta t$. For an interval of 20 MHz
(140-160 MHz) and an STEC=10~[tecu] we obtain for the difference in Faraday rotation
angles
$$ \Delta \beta_F=3\times 10^{8} /50 \times \Delta t=0.96$$
which is appreciable.

\end{appendix}

\end{document}